\documentclass[prc,aps,superscriptaddress,showpacs,nofootinbib,twocolumn]{revtex4}
\usepackage{amssymb}
\usepackage[dvips]{graphicx}
\usepackage[english]{babel}
\usepackage{indentfirst}
\usepackage{amsxtra}
\usepackage{amsmath}
\usepackage{supertabular}
\usepackage{multirow}
\usepackage[mathcal]{eucal}
\usepackage[usenames]{color}
\usepackage{ulem}

\begin{document}
\title {\textbf{A systematic study of giant quadrupole resonances with the subtracted second random--phase approximation: beyond--mean--field centroids and fragmentation}}
\author{O. Vasseur}
\affiliation{Institut de Physique Nucl\'eaire, CNRS-IN2P3, Universit\'e Paris-Sud,
Universit\'e Paris-Saclay, 91406 Orsay, France}
\author{D. Gambacurta}
\affiliation{Extreme Light Infrastructure - Nuclear Physics (ELI-NP), Horia Hulubei National Institute for Physics and Nuclear
Engineering, 30 Reactorului Street, RO-077125 M˘agurele, Jud. Ilfov, Romania}
\author{M. Grasso}
\affiliation{Institut de Physique Nucl\'eaire, CNRS-IN2P3, Universit\'e Paris-Sud,
Universit\'e Paris-Saclay, 91406 Orsay, France}

\begin{abstract}
A systematic analysis of giant quadrupole resonances is performed for several nuclei, 
from $^{30}$Si to $^{208}$Pb,  within the subtracted second random--phase--approximation (SSRPA) model
in the framework of the energy--density--functional theory.
Centroid energies and widths of the isoscalar giant quadrupole resonances are compared with the corresponding random--phase--approximation (RPA) values. 
We find lower SSRPA centroid energies compared to the RPA values leading, in general, to a 
better agreement with the experimental data. As far as the widths are concerned, we observe for both SSRPA and RPA cases a global 
 attenuation of the single--particle Landau damping going from lighter to heavier nuclei and 
we obtain, systematically, larger widths in the SSRPA model compared to the RPA case.  
For some selected nuclei for which high--resolution ($p,p'$) experimental data are available, namely $^{40}$Ca, $^{90}$Zr, $^{120}$Sn, and $^{208}$Pb,  
the theoretical strength distributions are directly compared with the experimental spectra. 
 We observe a significant improvement, with respect to RPA results, in the description of the spreading widths and of the fragmentation of the obtained spectra, due to the coupling between 1 particle-1 hole and 2 particle-2 hole configurations. 
\end{abstract}

\maketitle
\section{Introduction}
The emergence of collective excitations is one of the most interesting features 
of many-body systems. In atomic nuclei the most collective excitations are the so--called 
Giant Resonances (GRs) \cite{Speth,Harakeh} which are macroscopically interpreted as 
nuclear vibrations to which many nucleon take part coherently.  
 The random--phase--approximation (RPA) model provides a microscopic
description of the GRs constructed as  
superpositions of 1 particle–1 hole (1p1h) configurations. This approach is able 
to provide the gross features of GRs such as the centroid energy, the total strength  
and the corresponding energy-weighted sum rule (EWSR). However, other properties such as  
the GR’s fine structure, the damping mechanism and the decay width 
cannot be properly described in such a model based on the overlap of individual degrees of 
freedom.
 The total width of an excited mode is composed by three 
different contributions: i) the so--called Landau damping, corresponding to the 
fragmentation over 1p1h configurations; ii) the escape width due to the direct 
particle emission; iii) the spreading width, generated by the coupling between 
 1p1h configurations with, for instance, collective or multiparticle--multihole degrees of freedom. 
The interplay among these different contributions makes the description of the 
total width a very challenging task.  The Second RPA (SRPA) model is a natural extension of  RPA allowing for a more general description of the nuclear excitations and 
providing a valuable tool for the prediction of spreading widths and fine 
structure properties due to the introduction of 2 particle-2 hole (2p2h) configurations. 

A recent implementation of the SRPA model was illustrated in Refs. 
\cite{gamba2015,epja}, based on a subtraction procedure. This procedure, 
initially introduced for particle--vibration--coupling models \cite{tse1} and 
discussed more recently for extensions of the RPA \cite{tse2}, is designed to handle the problem of the 
double counting of correlations within 
energy--density--functional (EDF) theories. Such a double counting arises 
because the parameters of the effective interactions employed in EDF theories 
are adjusted in most cases to observables calculated at the 
mean--field level. The use of the same interactions in more 
sophisticated models intended to overcome the mean--field approximation may  
produce an overcounting of correlations, which is canceled by the subtraction 
procedure. In addition, such a procedure guarantees that the Thouless theorem is 
valid in any extensions of the RPA model \cite{tse2} and this is essential to 
ensure the validity of the stability conditions, related to the use of the 
Hartree-Fock ground state. Finally, the ultraviolet divergence occurring in all 
SRPA calculations done with zero--range forces and induced by the inclusion of 2p2h configurations is removed by the subtraction of the 
zero--energy self--energy \cite{gamba2015}. This means that all the drawbacks and the limitations of the SRPA model 
formulated in the EDF framework are  cured by the subtraction 
procedure, even if zero--range effective interactions are used. The subtracted 
SRPA (SSRPA) model thus represents a robust and stable theoretical tool 
for a beyond--mean--field description of 
the excitation spectra of many--body systems. 


Recently, we performed the first SSRPA calculations within a fully self--consistent scheme, by including in the residual interaction all the terms appearing in the effective interaction used for the description of the ground state \cite{plb2017}. The use of the subtraction procedure in a fully self--consistent scheme led to a satisfactory description of the low--lying dipole spectrum (below 10 MeV) and of the giant dipole resonance (GDR)  for the nucleus $^{48}$Ca \cite{plb2017}. A remarkable improvement was found in the low--energy spectrum with respect to previous SRPA calculations, where the subtraction procedure was not yet used and some terms were still missing in the residual interaction \cite{gambapygmy}, 
in the comparison with ($\gamma,\gamma'$) results of Ref. \cite{low}. Especially the values of the transition probabilities were strongly upgraded leading to a satisfactory agreement with the experimental data. In addition, an important improvement with respect to RPA calculations was found for the GDR of $^{48}$Ca, in the comparison with a recent measurement done at RCNP Osaka with the ($p,p'$) reaction 
at forward angle \cite{osaka}. The SSRPA model provided a much more realistic description of the spreading width of the GDR. 
The most important advantage of SRPA--based models, with respect to RPA, is indeed the possibility to describe the width and the fragmentation of the excited states, owing to the beyond--mean--field coupling between 1p1h and 2p2h configurations.  

In this article, we apply the SSRPA model based on Skyrme interactions to perform a systematic study of the centroids and the widths of the the isoscalar (IS) giant quadrupole resonance (GQR) for several nuclei, from medium--mass to heavy. In addition, we analyze the strength distributions for some selected nuclei for which high--energy resolution 
measurements are available. For these cases, we compare the theoretical 
predictions with 
RPA results and experimental spectra.

The article is organized as follows. We provide in Sec. \ref{forma} a brief overview of the SSRPA formalism. 
 We discuss in Sec. \ref{syste} the systematic trends obtained for the centroids and the widths in the SSRPA model. We illustrate detailed comparisons with RPA results for the centroids and the widths. Centroid energies are also compared with the available experimental data. We present in 
Sec. \ref{detail} a description of the strength distributions  for the nuclei $^{40}$Ca, 
$^{90}$Zr, $^{120}$Sn, and $^{208}$Pb, and we illustrate the comparison with RPA results and  
with experimental spectra. 
We show in Sec. \ref{cutoff} in an illustrative case the independence of the obtained results from the chosen energy cutoff in the 2p2h configurations. 
Conclusions are drawn in Sec. \ref{conclu}.

\section{Brief overview of the SSRPA model}
\label{forma}

All the details of the SRPA model implemented with a subtraction procedure may be found in Ref. \cite{gamba2015}. We provide here the main equations. It is well known that the SRPA equations can be put in the same compact form as the RPA equations, that is

\begin{equation}\label{eq_srpa}
\left(\begin{array}{cc}
  \mathcal{A} & \mathcal{B} \\
  -\mathcal{B}^{*} & -\mathcal{A}^{*} \\
\end{array}\right)
\left(%
\begin{array}{c}
  \mathcal{X}^{\nu} \\
  \mathcal{Y}^{\nu} \\
\end{array}%
\right)=\omega_{\nu}
\left(%
\begin{array}{c}
  \mathcal{X}^{\nu} \\
  \mathcal{Y}^{\nu} \\
\end{array}%
\right),
\end{equation}
where the matrices have different expressions compared to the RPA case, and the eigenvalues $\omega$ 
and eigenvectors ($\mathcal{X},\mathcal{Y}$) define the excitation energies and the wave functions of the excited states, respectively. Let us take as an illustration the matrix $\mathcal{A} $. In the standard SRPA model 
this matrix can be written as a block of matrices, 

\begin{displaymath}
\mathcal{A}=\left(\begin{array}{cc}
  A_{11'} & A_{12} \\
  A_{21} & A_{22'} \\
\end{array}\right),
\end{displaymath}
where '1' and '2' stand for 1p1h and 2p2h. The $A_{11'}$ matrix is the usual RPA matrix $A$. It is possible to write the SRPA equations as RPA--like equations with energy--dependent $A_{11'}$ and 
$B_{11'}$ matrices. In this case, the $A_{11'}(\omega)$ matrix reads 

\begin{widetext}

\begin{eqnarray}
A_{11^{\prime}} (\omega) = A_{11^{\prime}}+\sum_{2,2^{\prime}} A_{12} (\omega + i \eta - A_{22^{\prime}})^{-1} 
A_{2^{\prime}1^{\prime}} - 
\sum_{2,2^{\prime}} B_{12} (\omega + i \eta + A_{22^{\prime}})^{-1} 
B_{2^{\prime}1^{\prime}},
\label{arpa}
\end{eqnarray}

\end{widetext}
where the first term is the standard RPA matrix and the last one has to be included only in cases where density--depedent effective interactions are used (rearrangement terms) \cite{rearra}. Denoting by 
$E_{11'} (\omega)$ the energy--dependent part of $A_{11'} (\omega)$, the subtraction procedure 
proposed by Tselyaev consists in replacing $A_{11'}(\omega)$ by a subtracted matrix $A^S_{11^{\prime}} (\omega)$ written as 

\begin{equation}
A^S_{11^{\prime}} (\omega)= A_{11^{\prime}} (\omega) - E_{11^{\prime}}(0).
\label{sub1} 
\end{equation}

Coming back to an energy--independent form for the equations, the SSRPA matrix $\mathcal{A}$ read 

\begin{widetext}
\begin{eqnarray}
\mathcal{A}^S_D=\left(\begin{array}{cc}
  A_{11'}+ \sum_{22'} A_{12} (A_{22'})^{-1} A_{2'1'} + \sum_{22'} B_{12} (A_{22'})^{-1} B_{2'1'} & A_{12} \\
 & \\
  A_{21} & A_{22'} \\
\end{array}\right). 
\end{eqnarray}

\end{widetext}

In the calculations presented in this work such a matrix is fully computed and 
the diagonal approximation is not employed in the 2p2h sector $A_{22'}$, in spite of the huge 
numerical effort required to treat medium--mass and heavy nuclei. The subtractive term 
is instead calculated by using the diagonal approximation for the matrix $A_{22'}$ that has to be inverted. We showed in Ref. \cite{gamba2015}
that this approximation does not have a strong impact on the obtained excitation spectra allowing 
at the same time for a substantial reduction of the implied numerical effort.

\section{Systematic study for the IS GQR: centroid energies and spreading widths}
\label{syste}
From the experimental point of view, the IS GQR was extensively analyzed since its discovery 
more than 40 years ago \cite{pitthan,fukuda,lewis}. 
The first measurements were summarized 
in the 80s in a review on giant resonances \cite{ber} and in a 
systematic study dedicated to the giant monopole resonance (GMR) and to the GQR for several medium--mass and heavy nuclei, up to 
$^{208}$Pb \cite{you}. The IS GQR and GMR could be identified and distinguished one from the other using inelastic alpha scattering at small angles. Measurements for $^{48}$Ca \cite{lui}, $^{90}$Zr \cite{bor}, and Sn and Sm nuclei \cite{sha,li} are also available, based on inelastic scattering of alpha particles. 
Data taken on unstable nuclei were recently published: a measurement was first done on 
$^{56}$Ni, based on the reaction $^{56}$Ni$(d,d')$ \cite{mon} and, more recently, a measurement was perfomed on $^{68}$Ni using inelastic alpha and deuteron scattering \cite{van}. 

High--resolution experiments based on proton inelastic scattering have been performed at 
iThemba LABS 
to investigate the fine structure of GQR excitations for $^{40}$Ca \cite{usman}, $^{58}$Ni, $^{90}$Zr, $^{120}$Sn, and 
$^{208}$Pb \cite{she}. 

In this work, we extract the experimental data for the centroid energies from Ref. \cite{you} for almost all nuclei, with the exception 
of $^{48}$Ca \cite{lui}, $^{112}$Sn, $^{114}$Sn, $^{148}$Sm, $^{150}$Sm, $^{152}$Sm \cite{sha}. Data on the IS GQR are also available for Sn isotopes in Ref. \cite{li}. The centroid energies of Ref. \cite{li} are in rather good agreement with the values obtained in other measurements. On the other side, the widths reported in Ref. \cite{li} are much larger (the double) compared to those obtained with other measurements \cite{you,sha}. Due to these ambiguities, we decided to show a systematic comparison with the experimental data only for the centroid energies. As far as width, fine structure and fragmentation are concerned, 
we dedicate a more focused discussion in Section \ref{detail} where we select only nuclei for which high--precision ($p,p'$) data are available. A comparison between theoretical and experimental results is done only for these selected cases. 

We analyze thirteen spherical--expected medium--mass and heavy nuclei: $^{30}$Si, $^{34}$Si, $^{36}$S, $^{40}$Ca, 
$^{48}$Ca, $^{56}$Ni, $^{68}$Ni, $^{90}$Zr, $^{114}$Sn, $^{116}$Sn, $^{120}$Sn, $^{132}$Sn, and $^{208}$Pb. 
We perform RPA and SSRPA calculations with the SLy4 parametrization \cite{sly4} of the Skyrme interaction.
The single--particle space is chosen large enough to assure that the  EWSR are preserved within 1\%.
For the 2p2h space in the SSRPA calculations, we
use a cutoff of 60 MeV for medium--mass nuclei ($^{30}$Si, $^{34}$Si, $^{36}$S, $^{40}$Ca, 
$^{48}$Ca, $^{56}$Ni, $^{68}$Ni) and of 50 MeV for the heavy ones ($^{90}$Zr, $^{114}$Sn, $^{116}$Sn, $^{120}$Sn, $^{132}$Sn, and $^{208}$Pb). We checked that these cutoff values provide stable results.

Centroid energies $E_c$ and widths $\Gamma$ are usually estimated by using the moments of the strength 
 $m_0$, $m_1$, and $m_2$, namely 
\begin{equation}
E_c=\frac{m_1}{m_0} 
\label{ec}
\end{equation}  
and 
\begin{equation}
\Gamma=\sqrt{m_2/m_0-(m_1/m_0)^2}.
\label{gam}
\end{equation}  
However, this estimation turns out to be reasonable only in those cases where the strength is well concentrated around a main peak (which is typically the case for RPA calculations done for spherical nuclei) and when the spectrum contains more than a single dominant peak. Now, it turns out that in SRPA--based models the strength may be strongly fragmented. In these cases, such a procedure for extracting the centroids and the widths may 
alter the results reducing  
 in an artificial way the estimated value for the width. 
On the other side, also for those RPA spectra where there is only a single significant peak this procedure is not adequate, the resulting width being artificially too large. 

To make more realistic estimations,  
centroids and widths of the IS GQRs were computed here in a similar way as in Ref. \cite{sl}, that is by fitting a Lorentzian distribution. This adjustement was done on curves obtained by folding the discrete spectra with narrow Lorentzian distributions. 
In the SSRPA case, the folding done with very narrow distributions follows fairly well the extremely 
dense SSRPA spectra and no artificial effects induced by the performed folding are observed when the Lorentzian distribution is fitted. On the other side, in those RPA cases where there is a unique dominant peak, the width extracted in this way turns out to be equal to the width of the folding Lorentzian. We used in these cases very narrow folding Lorentzian (100--keV width) not to induce any artificial spreading effect.

\begin{widetext}

\begin{figure}
\includegraphics[scale=0.45]{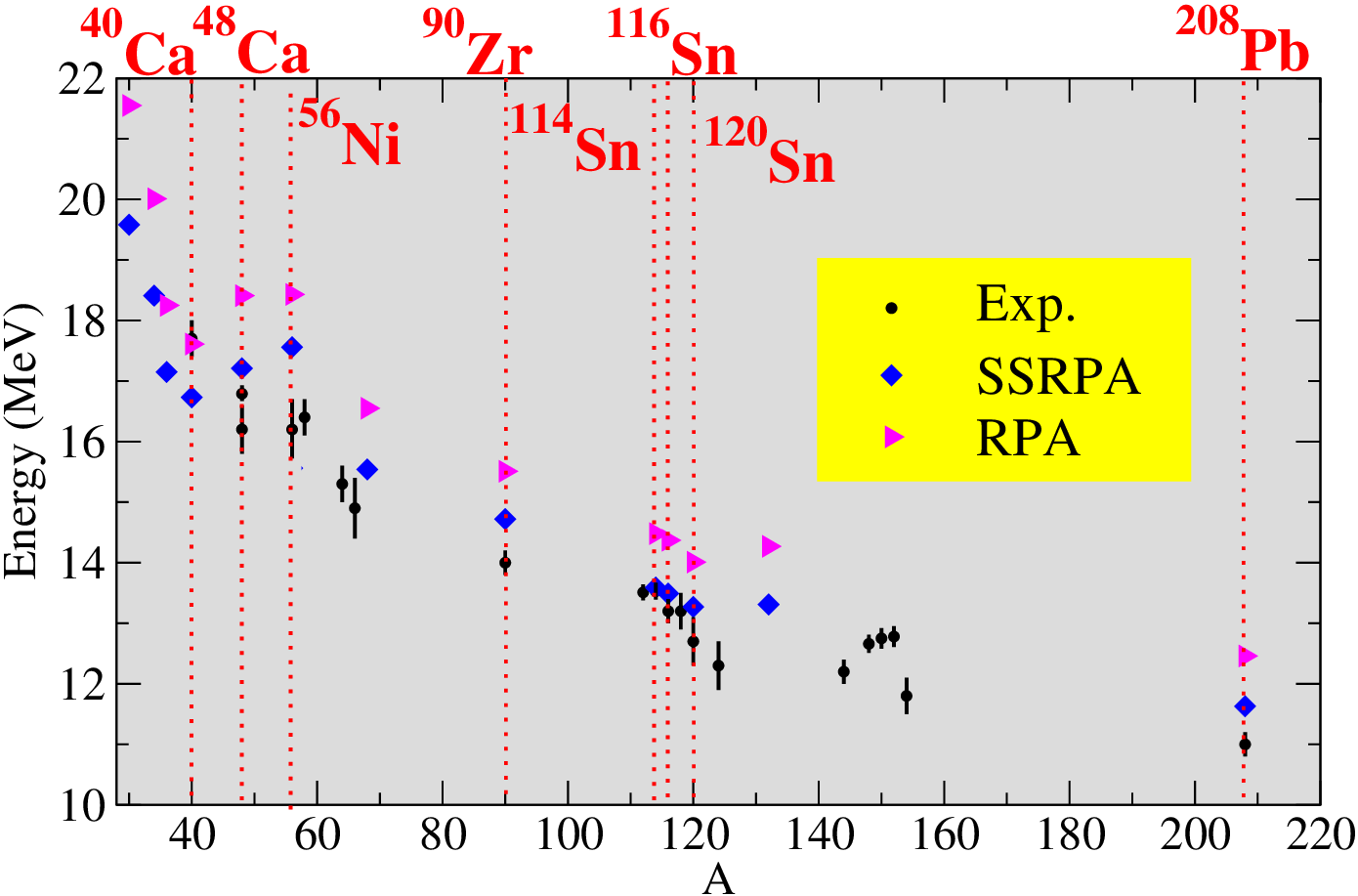}
\caption{Centroids of the IS GQR. The experimental data are displayed as black circles (with their associated error bars) and are extracted from Refs. \cite{you,lui,sha,mon}. SSRPA (RPA) predictions are plotted as blue diamonds (magenta triangles). At $A=48$, there are two experimental measurements, for $^{48}$Ti and for $^{48}$Ca. The experimental point corresponding to $^{48}$Ca is the highest one. Theoretical calculations are performed for the nuclei $^{30}$Si, $^{34}$Si, $^{36}$S, $^{40}$Ca, $^{48}$Ca, $^{56}$Ni, $^{68}$Ni, $^{90}$Zr, $^{114}$Sn, $^{116}$Sn, $^{120}$Sn, $^{132}$Sn, and $^{208}$Pb. }
\label{fig1}
\end{figure}

\begin{figure}
\includegraphics[scale=0.45]{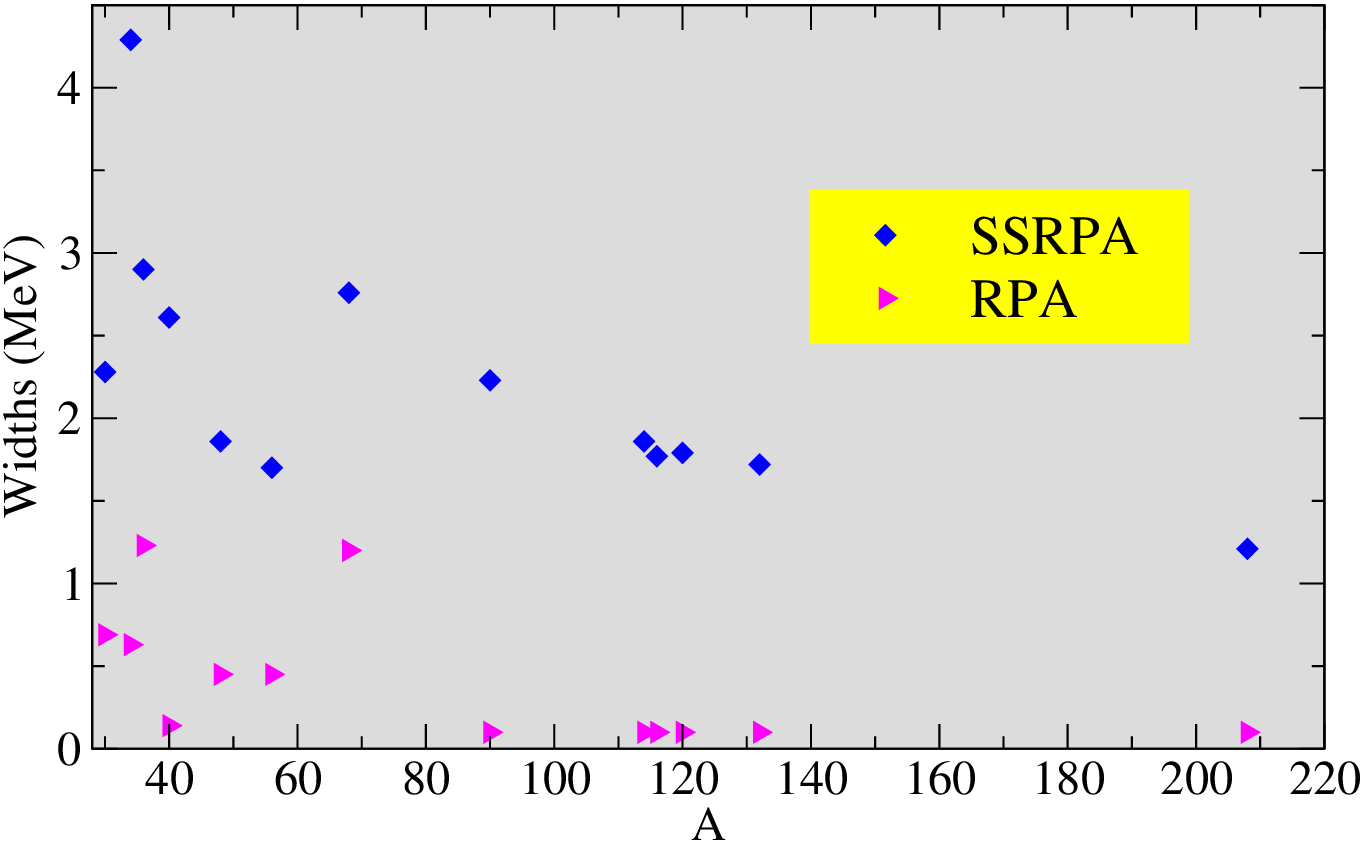}
\caption{Theoretical widths calculated with the fit of a Lorentzian distribution within RPA and SSRPA models.}
\label{fig2}
\end{figure}

\end{widetext}


The systematic trend provided by the SSRPA model for the centroids is shown in Fig. \ref{fig1} where  SSRPA results (blue diamonds) are compared with the corresponding RPA results (magenta triangles) and with the experimental data (black circles) represented in the figure with the
corresponding error bars. 
Nuclei for which a comparison between our theoretical predictions and the corresponding experimental data may be done are 
identified in the figure by vertical dotted red lines.  
We observe that the SSRPA centroids are systematically located  at lower energies than the RPA values. 
It is known that the centroid energies of the IS GQR are strongly related to the effective mass (see for example two recent reviews, Refs. \cite{bao,roca}).
A discussion on this aspect and on the related impact on the modification of the effective mass beyond the mean--field approximation   
is done in a dedicated work  \cite{grasso18}.

\begin{figure}
\includegraphics[scale=0.35]{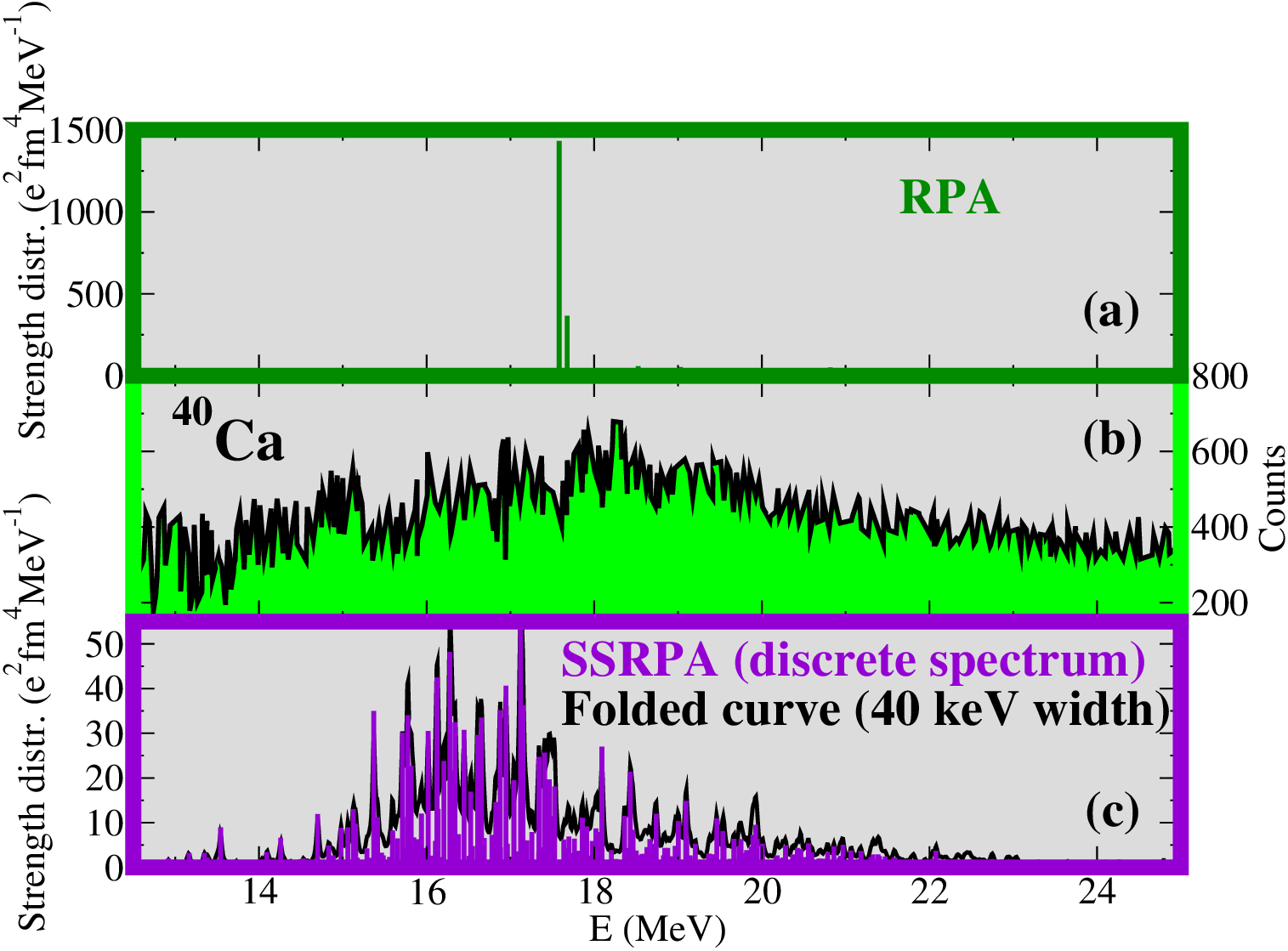}
\caption{(a) RPA strength distributions calculated for the nucleus $^{40}$Ca; (b) Experimental spectrum \cite{usman} for the IS GQR for $^{40}$Ca; (c) SSRPA strength distributions calculated for the nucleus $^{40}$Ca.}
\label{figca}
\end{figure}

\begin{figure}
\includegraphics[scale=0.35]{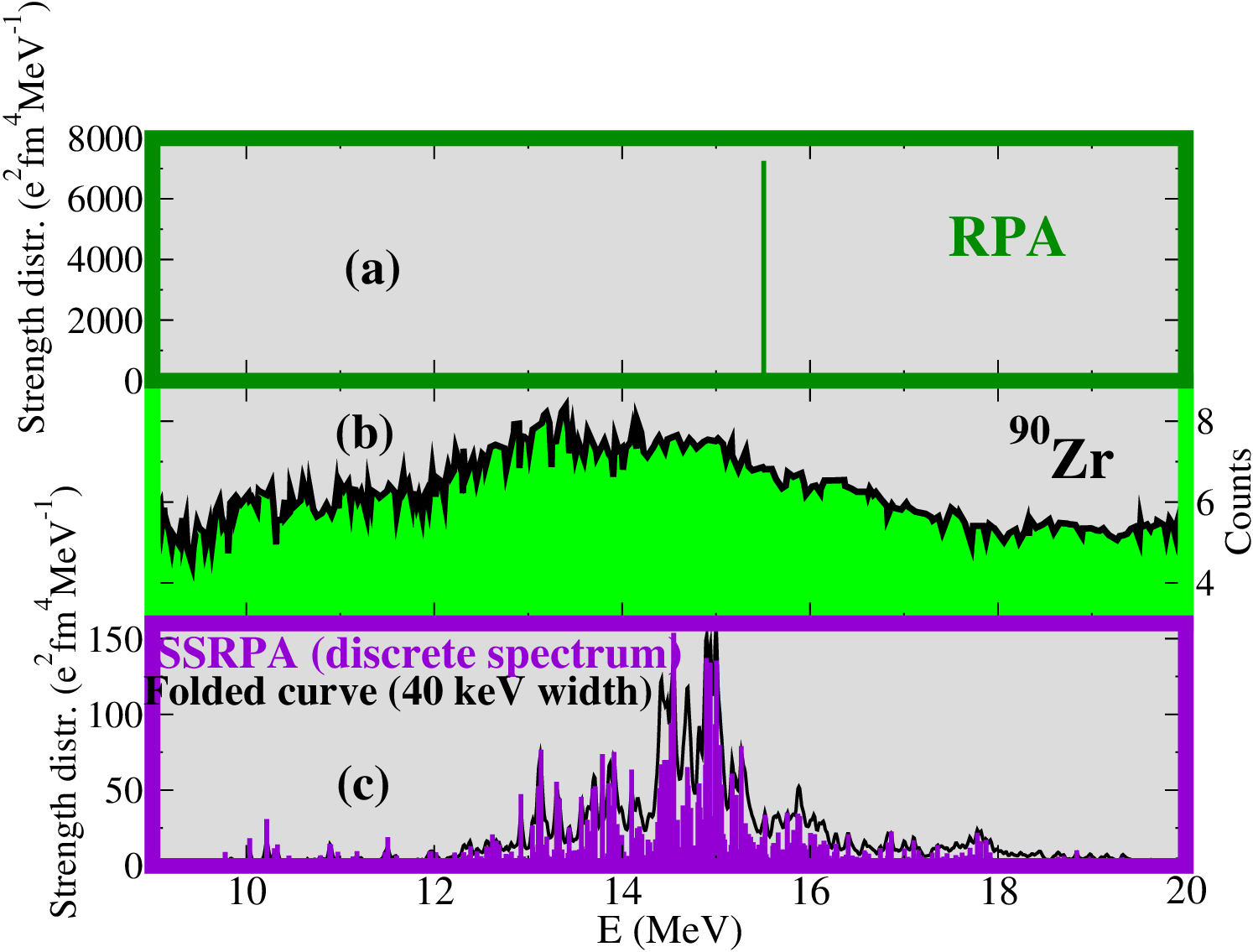}
\caption{Same as in Fig. \ref{figca} but for the nucleus $^{90}$Zr. The experimental data are this time extracted from Ref. \cite{she}.}
\label{figzr}
\end{figure}

For most of the cases where the experimental centroids are available, we observe that the SSRPA energies are in better agreement with the experimental values than the RPA centroids, which in general overestimate the data, as can be seen in Fig. \ref{fig1}.

We expect that the description of the widths is strongly modified in the SSRPA model, compared to the RPA case, because an additional spreading effect is explicitly taken into account (in addition to the single--particle Landau damping which is already present in RPA) owing to the coupling between 1p1h and 2p2h configurations. 
Figure \ref{fig2} displays the SSRPA and RPA widths. 
We observe that the SSRPA widths are, as expected, systematically larger than the RPA ones. 
Figure \ref{fig2} indicates also another interesting trend: 
 globally, both in RPA and in SSRPA, the widths are reduced going from lighter to heavier nuclei implying that there is a more important fragmentation in lighter than in heavier nuclei. Since this effect is observed already at the RPA level, we deduce that the higher fragmentation for lighter nuclei is produced by a stronger Landau damping, which is an effect taken into account both in RPA and in SSRPA models. The importance of such a single--particle fine--structure effect in lighter nuclei was already discussed in Refs. \cite{usman,aiba1,aiba2}. In particular, the authors of Refs. \cite{aiba1,aiba2} compared the cases of $^{40}$Ca and $^{208}$Pb and illustrated the differences in the damping mechanism arising for the two nuclei of different mass, the lighter one being more affected by the single--particle Landau damping.  
Although this trend indicating a Landau--damping attenuation is observed globally also in our results, we notice that, for the nucleus $^{40}$Ca, we do not find any important effect related to the Landau damping, the RPA width being particularly small in this case. For this nucleus, the beyond--mean--field effects coming from the mixing with 2p2h configurations are particularly important and produce a strong increase of the width going from RPA to SSRPA. 

\begin{figure}
\includegraphics[scale=0.35]{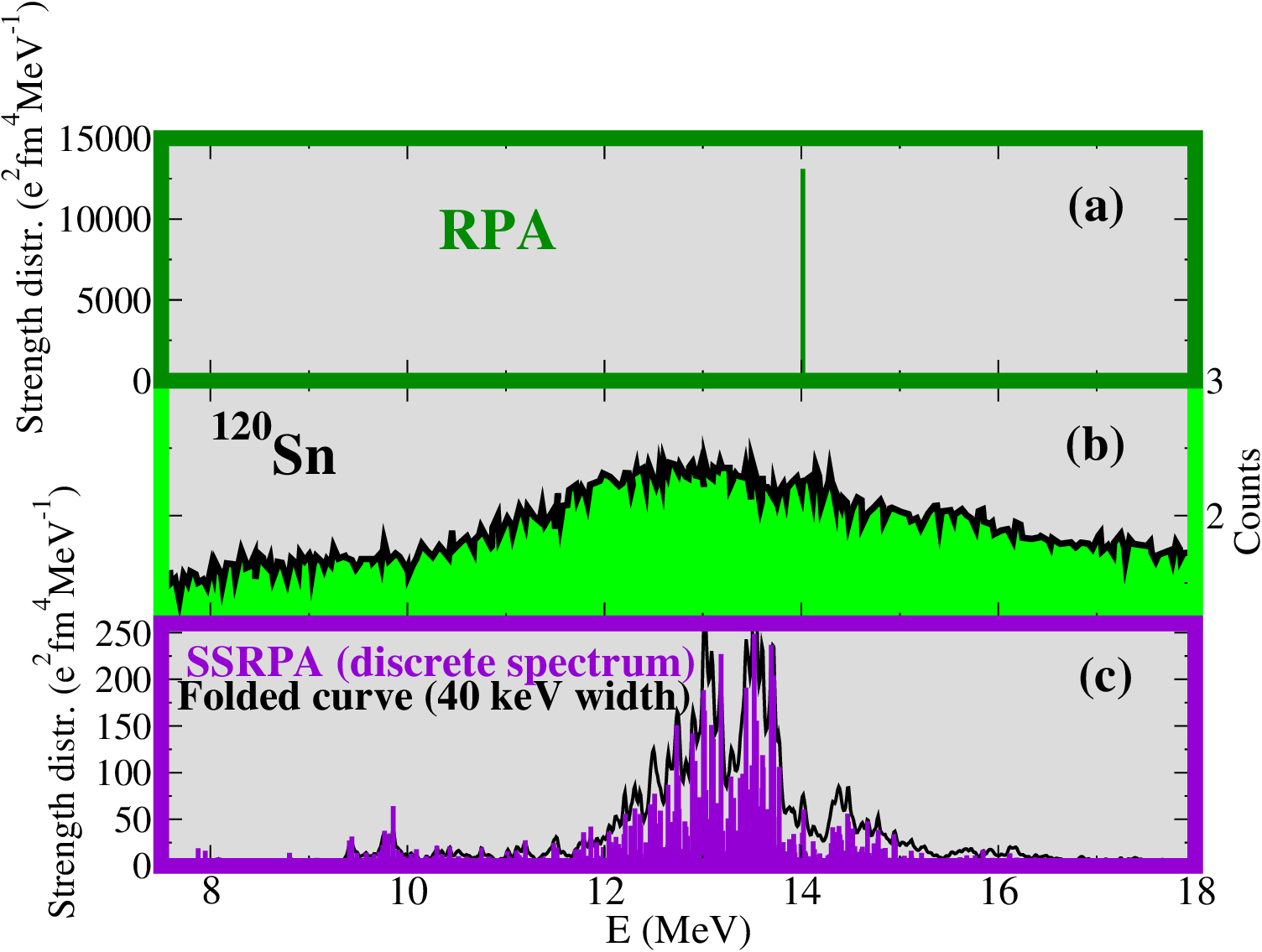}
\caption{Same as in Fig \ref{figzr} but for the nucleus $^{120}$Sn.}
\label{figsn}
\end{figure}

%

\section{Detailed analysis for$^{40}$Ca, $^{90}$Zr, $^{120}$Sn, and $^{208}$Pb } 
\label{detail}
High--resolution ($p,p'$) spectra are available for the IS GQRs of the nuclei $^{40}$Ca, $^{90}$Zr, 
$^{120}$Sn, and $^{208}$Pb. Energy resolutions of $\sim$ 40 keV could be achieved and the fine structure of the excitation spectra could be examined.

\begin{figure}
\includegraphics[scale=0.35]{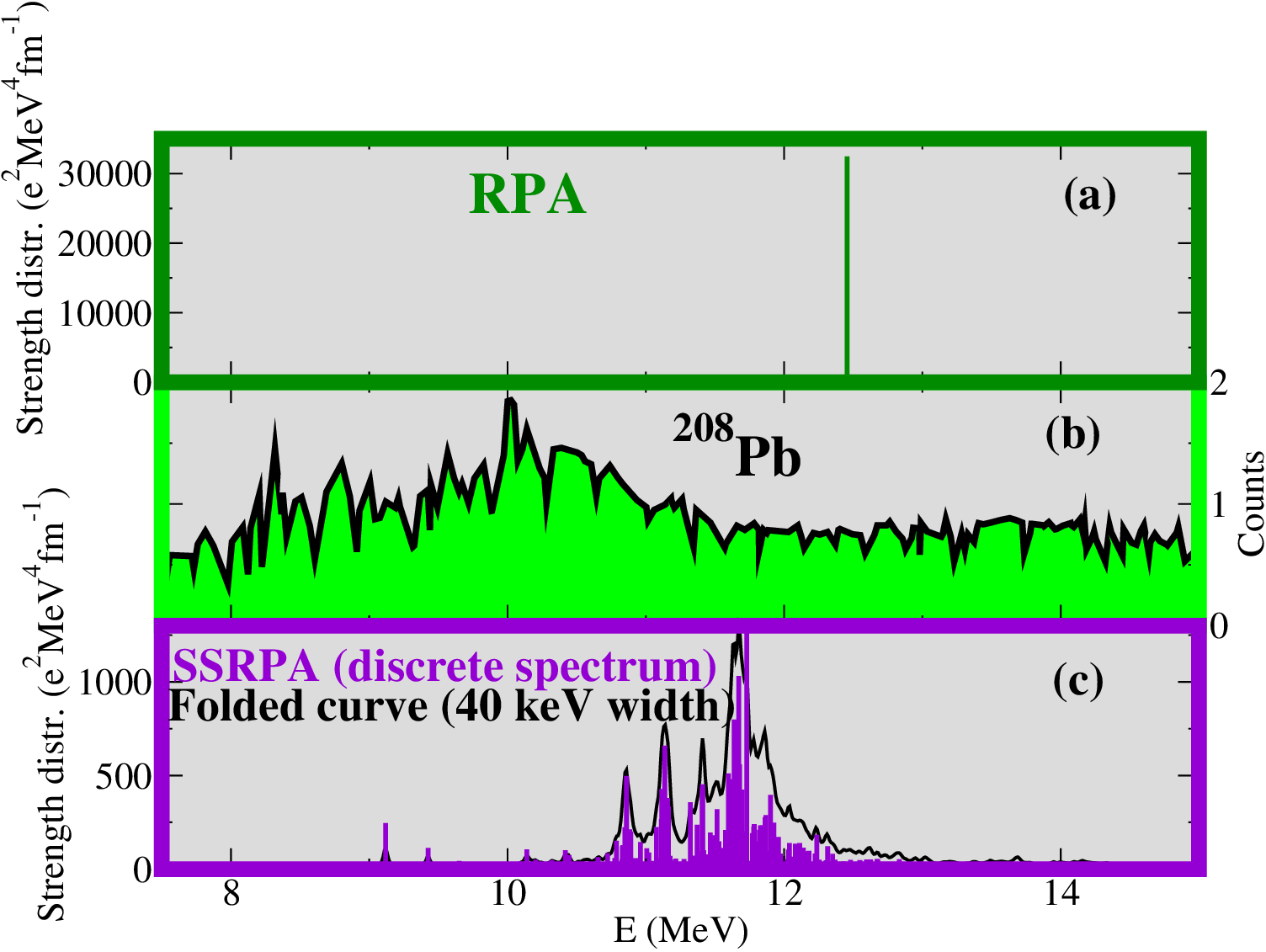}
\caption{Same as in Fig. \ref{figsn} but for the nucleus $^{208}$Pb.}
\label{figpb}
\end{figure}

We present in Fig. \ref{figca}(c) the SSRPA strength distribution (violet bars) for the nucleus 
$^{40}$Ca. To better compare it with the corresponding experimental spectrum (b), a folded curve is also plotted (black solid line and grey area), obtained by folding the discrete distribution with a Lorentzian of width equal to 40 keV, which corresponds to the experimental energy resolution. We observe that the folded curve follows well the fine structure provided by the discrete spectrum. The RPA strength distribution 
is also shwon in Fig. \ref{figca}(a). A single significant peak is found in this case. 
This is the only case shown in Fig. \ref{fig1} where the SSRPA centroid energy is slightly underestimated compared to the experimental value. The RPA centroid is in better agreement with it. However, the significant advantage of using the SSRPA model instead of RPA is clearly indicated by Fig. \ref{figca}. Our RPA prediction displays a unique dominant peak, whereas the SSRPA strength distribution is much more fragmented and extends over a larger energy region where the experimental data are spread. Figure \ref{figzr} shows the same quantities as in Fig. \ref{figca} but for the nucleus $^{90}$Zr. In this case, the RPA centroid is larger by more than 1 MeV compared to the experimental value (Fig. \ref{fig1}). The SSRPA prediction is located at lower energies, in better agreement with data. Again, a relevant improvement with respect to RPA is observed in the strength fragmentation (unique dominant peak in RPA). The same comments bay be extended to Figs. \ref{figsn} and \ref{figpb} where results for $^{120}$Sn and $^{208}$Pb are presented. 

In general, all the SSRPA centroids are slightly shifted downwards compared to RPA. In almost all cases (with the exception of $^{40}$Ca) this leads to a better agreement with experimental data. As far as fragmentation and fine structure are concerned, we observe a substantial improvement in the SSRPA results compared to RPA where, for the four cases under consideration here, the strength distribution is characterized by a single dominant peak. The SSRPA strength distribution is spread over a much larger window, where the experimental response is located. 
We observe that the comparison with the experimental fine structure shows a qualitative global agreement in the sense that our model provides a fragmented response in the same energy region. 
We note however that, in all cases, the energy window where the experimental strength is distributed 
is broader than the range where the SSRPA response is located. This is probably related to missing effects in our theoretical model, such as the inclusion of higher--order configurations (3 particles-3 holes,...) and of spreading effects induced by the coupling with the continuum, not taken into account here. 

We also notice that our theoretical predictions are qualitatively and quantitatively different from those published in Ref. \cite{usman} and compared to the high--precision spectrum of $^{40}$Ca. Those theoretical results (both RPA and SRPA) are based on a potential derived from a realistic interaction with the Unitary Correlation Operator Method (UCOM) and provide: (i) a more fragmented (than ours) RPA spectrum; (ii) a strongly overestimated RPA centroid; (iii) a SRPA spectrum strongly shifted downwards with respect to RPA and not corrected from instabilities; (IV) a SRPA spectrum which is much less dense than ours in the strength distribution. These differences (apart from the fact that, in any case, no subtraction procedures are used in SRPA calculations of Ref. \cite{usman}) may probably be ascribed to the use of a potential derived from a realistic interaction (Argonne V18) which generates, at the Hartree-Fock level, a single--particle spectrum with very large interlevel spacings. This is the reason why the RPA centroid is located so high in energy. And this is probably also the reason why the coupling with 2p2h configurations in the SRPA model is not able in that case to produce a dense strength distribution, in spite of the huge number of elementary 2p2h configurations.



\section{Stability of the results with respect to the energy cutoff}
\label{cutoff}

The subtraction procedure was not designed to remove the 
ultraviolet divergence generated by the use of zero--range forces in SRPA--based models.
Nevertheless, the stability of the obtained results with respect to the energy cutoff in the 2p2h configurations was already noticed in Ref. \cite{gamba2015}. It turns out that the subtraction of the zero--energy self--energy removes the divergent contribution leading to cutoff--independent results. 

\begin{figure}
\includegraphics[scale=0.36]{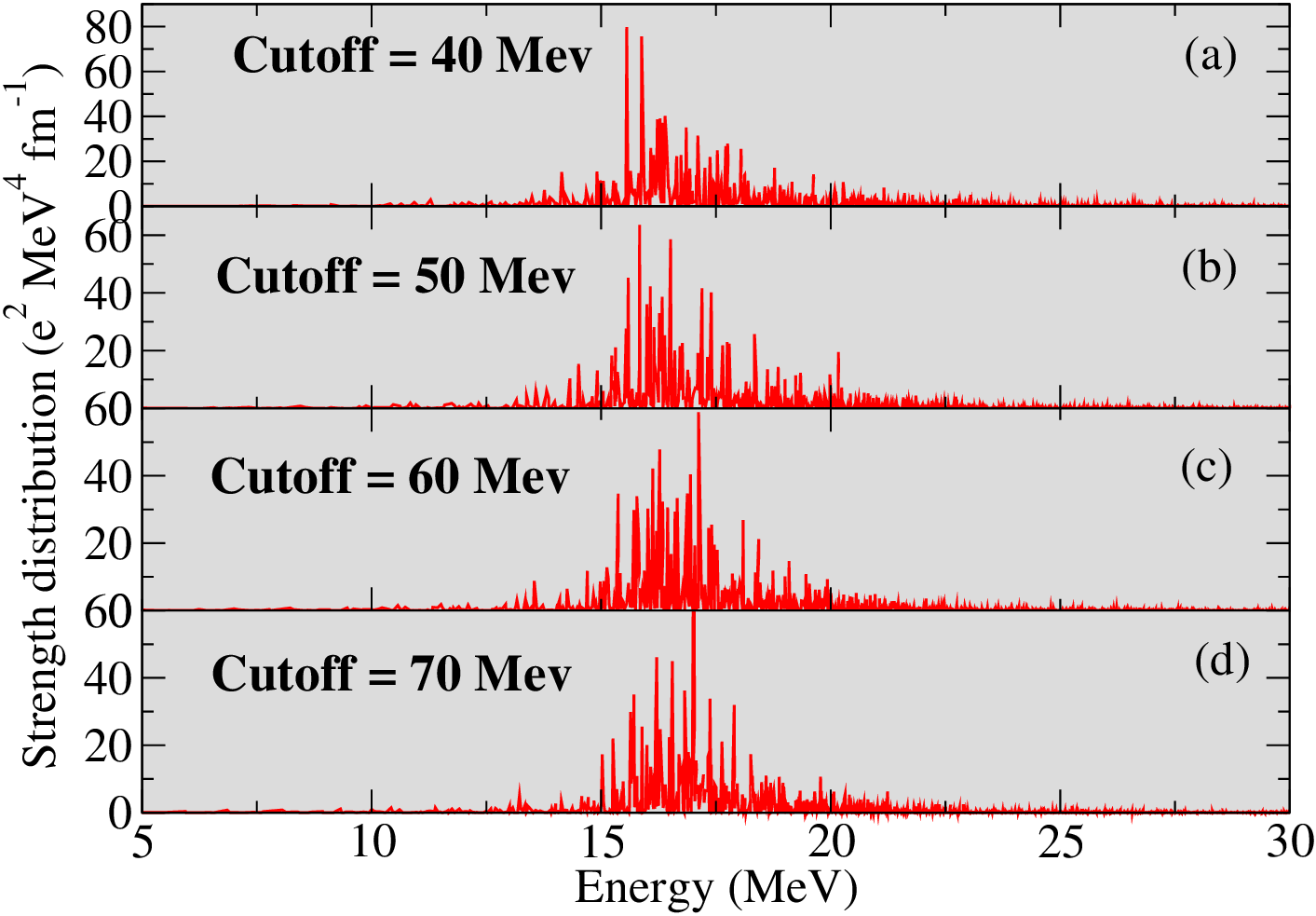}
\caption{Strength distributions obtained for the nucleus $^{40}$Ca with four different cutoff values for the 2p2h configurations, 40 (a), 50 (b), 60 (c), and 70 (d) MeV.}
\label{cut}
\end{figure}

To further underline the robustness of the SSRPA model in this respect we show in Fig. \ref{cut} 
an illustrative case. Figure \ref{cut} displays the discrete strength distributions obtained for the 
nucleus $^{40}$Ca with four cutoff values in the 2p2h configutations, from 40 to 70 MeV. 
We definitely notice that the independence of the results from the cutoff is achieved. 
Centroid energies are stable. For instance, going from 60 to 70 MeV the centroid energy is shifted from 16.74 to 16.86 MeV. This extremely weak modification of the centroid, less than 1 \%, indicates that the 
cutoff dependence induced by the ultraviolet divergence is removed by the subtractive term.

\section{Conclusions}
\label{conclu}

IS GQRs are studied using the SSRPA model in the framework of EDF theories for 
thirteen spherical--expected medium--mass and heavy nuclei. The SSRPA 
predictions for the centroids are compared with RPA results and experimental 
data. The comparison between the  SSRPA and RPA widths is also presented ad discussed. 

SSRPA centroids are slightly shifted downwards with respect to RPA values and are, 
globally, in better agreement with experimental data. SSRPA widths are 
systematically larger than the RPA ones, as expected. 
An attenuation of the single--particle Landau damping is also observed both in 
SSRPA and RPA results going from medium--mass to heavier nuclei. 

For $^{40}$Ca, $^{90}$Zr, $^{120}$Sn, and $^{208}$Pb the theoretical strength 
distributions are  compared with the experimental response. For these nuclei, 
high--resolution ($p,p'$) experimental data are available. A significant 
improvement of the description of the spreading width is found in SSRPA compared 
to RPA (where a single dominant peak is predicted) and fragmented strengths are 
obtained. However, they do not extend over the whole energy region where the 
experimental data are located. This is probably due to missing effects such as 
highr--order correlations or continuum coupling. In spite of this, a clear 
important improvement with respect to the mean--field based RPA model is found. 

%


\begin{thebibliography}{10}
\bibitem{Speth}  J. Speth and J. Wambach, \textit{Theory of giant resonance, in: Electric and Magnetic Giant
Resonances in Nuclei}, in: Int. Rev. Nucl. Phys., vol. 7, World Scientific, 1991, p. 1.
\bibitem{Harakeh}  M. N. Harakeh and A. Van der Woude, \textit{Giant Resonances}
(Clarendon, Oxford, UK, 2001).

\bibitem{gamba2015} D. Gambacurta, M. Grasso, and J. Engel, Phys. Rev. C (2015).
\bibitem{epja} D. Gambacurta and M. Grasso, Eur. Phys. J. A 52, 198 (2016).
\bibitem{tse1} V.I. Tselyaev, Phys. Rev. C 75, 024306 (2007).
\bibitem{tse2} V.I. Tselyaev, Phys. Rev. C 88, 054301 (2013).
\bibitem{plb2017} D. Gambacurta, M. Grasso, and O. Vasseur, Phys. Lett. B 777, 163 (2018).  
\bibitem{gambapygmy} D. Gambacurta, M. Grasso, and F. Catara, Phys. Rev. C 84, 034301 (2011).
\bibitem{low} T. Hartmann, J. Enders, P. Mohr, K. Vogt, S. Volz, and A. Zilges, Phys. Rev. C 65, 034301 (2002).
\bibitem{osaka} J. Birkhan, et al., Phys. Rev. Lett. 118, 252501 (2017).
\bibitem{rearra} D. Gambacurta, M. Grasso, and F. Catara, J. Phys. G: Nucl. and Part. Phys. 38, 035103 (2011).
\bibitem{pitthan} R. Pitthan and T. Walcher, Phys. Lett. B 36, 563 (1971).
\bibitem{fukuda} S. Fukuda and Y. Torizuka, Phys. Rev. Lett. 29, 1109 (1972).
\bibitem{lewis} M. Lewis and F. Bertrand, Nucl. Phys. A 196, 337 (1972).
\bibitem{ber} F.E. Bertrand, Nucl. Phys. A 354, 129c (1981).
\bibitem{you} D.H. Youngblood, P. Bogucki, J.D. Bronson, U. Garg, Y.-W. Lui, and M Rozsa, Phys. Rev. C 23, 1997 (1981).
\bibitem{lui} Y.-W. Lui, D.H. Youngblood, S. Shlomo, X. Chen, Y. Tokimoto, Krishichayan, M. Anders, and J. Button, Phys. Rev. C 83, 044327 (2011). 
\bibitem{bor} W.T.A. Borghols, et al., Nucl. Phys. A 504, 231 (1989). 
\bibitem{sha} M.M. Sharma, W.T.A. Borghols, S. Brandenburg, S. Crona, A. van der Woude, and M.N. Harakeh, Phys. Rev. C 38, 2562 (1988).
\bibitem{li} T. Li et al., Phys. Rev. C 81, 034309 (2010).
\bibitem{mon} C. Monrozeau, et al., Phys. Rev. Lett. 100, 042501 (2008).
\bibitem{van} M. Vandebrouck, et al., Phys. Rev. C 92, 024316 (2015). 
\bibitem{usman} I. Usman, et al., Phys. Lett. B 698, 191 (2011).
\bibitem{she} A. Shevchenko, et al., Phys. Rev. Lett. 93, 122501 (2004).
\bibitem{sly4} E. Chabanat, P. Bonche, P. Haensel, J. Meyer, R. Schaeffer, Nucl. Phys. A 627, 710 (1997); {\it{ibid}} A 635, 231 (1998); {\it{ibid}} 643, 441 (1998). 
\bibitem{sl} G. Scamps, D. Lacroix, Phys. Rev. C 88, 044310 (2013). 
\bibitem{bao} Bao-An Li, Bao-Jun Cai, Lie-wen Chen, Jun Xu, Prog. Part. Nucl. Phys. 99, 29 (2018).
\bibitem{roca} X. Roca-Maza and N. Paar, arXiv:1804/0625v1.
\bibitem{grasso18} M. Grasso, D. Gambacurta, O. Vasseur, in progress
\bibitem{aiba1} H. Aiba, M. Matsuo, S. Nishizaki, and T. Suzuki, Phys. Rev. C 68, 054316 (2003).
\bibitem{aiba2} H. Aiba, M. Matsuo, S. Nishizaki, and T. Suzuki, Phys. Rev. C 83, 024314 (2011).


\end{thebibliography}
\end{document}